\documentclass[
    ,final            % use final for the camera ready runs
,sort&compress
  ]
  {aipproc}

\layoutstyle{8x11single}
\usepackage[utf8]{inputenc}

\usepackage{amsmath}
\usepackage{amssymb}

\DeclareMathOperator{\Tr}{Tr}

\begin{document}

\title{Hadron Structure from Lattice QCD}


\author{Jeremy Green}{
  address={Institut für Kernphysik, Johannes Gutenberg-Universität Mainz, D-55099 Mainz, Germany},
  email={green@kph.uni-mainz.de}
}

\begin{abstract}
  Recent progress in lattice QCD calculations of nucleon structure
  will be presented. Calculations of nucleon matrix elements and form
  factors have long been difficult to reconcile with experiment, but
  with advances in both methodology and computing resources, this
  situation is improving.  Some calculations have produced agreement
  with experiment for key observables such as the axial charge and
  electromagnetic form factors, and the improved understanding of
  systematic errors will help to increase confidence in predictions of
  unmeasured quantities. The long-omitted disconnected contributions
  are now seeing considerable attention and some recent calculations
  of them will be discussed.
\end{abstract}

\maketitle

\section{Introduction}

For many observables, lattice QCD calculations have achieved careful
control over systematic errors and obtained precise
results~\cite{Aoki:2013ldr}. However, for many others, this is still a
work in progress, and control over systematics remains a
challenge. The latter holds true for nucleon structure --- which is
the focus of this review --- although much progress is being made.

Lattice regularizations of QCD use a discretized Euclidean hypercubic
spacetime lattice, usually with a spacing $a$ between points and a
finite periodic box with dimensions $L_s^3\times L_t$. The path
integral is performed analytically over quark fields and numerically
using Monte Carlo methods over the gauge fields, so that in practice a
calculation has two separate steps: First, an \emph{ensemble} of
\emph{gauge configurations} $\{U\}$ is produced by sampling the path
integral. Second, observables are computed on each gauge
configuration. For observables involving quark fields, this requires
computing parts of the quark propagator $S[U]=D^{-1}[U]$ on each
background gauge configuration, where $D[U]$ is the Dirac operator. As
efficient algorithms are available for solving $D[U]\psi=\eta$,
observables that can be constructed using source-to-all propagators
typically require the least computational resources.

To determine nucleon matrix elements $\langle
p'|\mathcal{O}|p\rangle$, we compute two-point and three-point
functions,
\begin{align}
  C_\text{2pt}(\vec p,t) &= \sum_{\vec x}e^{-i\vec p\cdot(\vec x-\vec x_0)} \Tr[\Gamma_\text{pol}\langle N(x,t_0+t) \bar N(x_0,t_0) \rangle], \label{eq:c2pt}\\
C_\text{3pt}^\mathcal{O}(\vec p',\vec p,\tau,T) &= \sum_{\vec x,\vec y}e^{-i\vec p'\cdot(\vec x-\vec x_0)}e^{i(\vec p'-\vec p)\cdot(\vec y-\vec x_0)} \Tr[\Gamma_\text{pol}\langle N(x,t_0+T) \mathcal{O}(y,t_0+\tau) \bar N(x_0,t_0) \rangle],
\end{align}
where $N$ is a nucleon interpolating operator and $\Gamma_\text{pol}$
is a polarization matrix. Typically, $N=\epsilon^{abc}(\tilde
u_a^TC\gamma_5\tilde d_b)\tilde u_c$, where $\tilde q$ is a spatially
``smeared'' quark field with spatial extent tuned to optimize the
overlap of $N$ with the ground state, and $\mathcal{O}$ is an operator
bilinear in quark fields.

The Wick contractions for the three-point functions include the
notorious \emph{disconnected diagrams}, which require the quark
propagator from every point on the operator $\mathcal{O}$'s timeslice
back to itself and would thus be very expensive to compute exactly;
recent calculations using stochastic estimation for these disconnected
loops will be discussed later in this review. Because of the
difficulty in computing disconnected diagrams, much of the focus in
lattice calculations has been on isovector observables, which have no
disconnected contribution.

The other three-point-function contractions form the \emph{connected
  diagrams}, which can be computed exactly (for each gauge
configuration and source position) using the sequential propagator
method, where a propagator from the source point $(x_0,t_0)$ is used
to create a ``source'' on timeslice $t_0+T$ for a second
propagator. The forward and sequential propagators can then be
combined using any quark bilinear operator at any point on the
lattice. However, changing the parameters of the annihilation
operator, such as the source-sink separation $T$ or the momentum $\vec
p'$, requires computing a new sequential propagator.

In addition to the ground-state nucleon, the interpolating operator
will couple to other states with the same quantum numbers. When all
time separations $t$, $\tau$, and $T-\tau$ become large, excited
nucleon states decay more rapidly than the ground state and the
dominant contribution to the two-point and three-point functions comes
from the ground-state nucleon. Specifically,
\begin{align}
C_\text{2pt}(\vec p,t) &\to Z(\vec p)^2 e^{-E(\vec p)t},\\
C_\text{3pt}^\mathcal{O}(\vec p',\vec p,\tau,T) &\to Z(\vec p')Z(\vec p) e^{-E(\vec p')(T-\tau)} e^{-E(\vec p)\tau} \langle p'|\mathcal{O}|p\rangle,
\end{align}
where $Z(\vec p)$ is an overlap factor between the interpolating
operator and the ground state. The traditional method for determining
the matrix element is to construct a \emph{ratio} to cancel the
overlap factors and the time dependence:
\begin{equation}
\begin{aligned}
R^\mathcal{O}(\vec p',\vec p,\tau,T) &= \frac{C_\text{3pt}^\mathcal{O}(\vec p',\vec p,\tau,t)}{\sqrt{C_\text{2pt}(\vec p,T)C_\text{2pt}(\vec p',T)}}\sqrt{\frac{C_\text{2pt}(\vec p,T-\tau)C_\text{2pt}(\vec p',\tau)}{C_\text{2pt}(\vec p',T-\tau)C_\text{2pt}(\vec p,\tau)}}\\
&= \langle p'|\mathcal{O}|p\rangle + O(e^{-\Delta E(\vec p)\tau}) + O(e^{-\Delta E(\vec p')(T-\tau)}),
\end{aligned}
\end{equation}
where $\Delta E(\vec p)$ is the energy gap to the lowest-lying excited
state with momentum $\vec p$. For each source-sink separation $T$,
choosing the midpoint $\tau=T/2$ yields leading excited-state
contaminants that decay as $e^{-\Delta E_\text{min}T/2}$, where
$\Delta E_\text{min} = \text{min}\{\Delta E(\vec p),\Delta E(\vec
p')\}$.

To eliminate excited states, we want to use large source-sink
separations $T$; however, the signal-to-noise ratio decays rapidly,
with an asymptotic behaviour $\sim
e^{-(m_N-3m_\pi/2)T}$~\cite{Lepage:1989hd}. It is thus
challenging to use large-enough source-sink separations that
excited-state effects are negligible, while still obtaining a good
signal. This is especially true at smaller pion masses, since the
signal decays more rapidly and the lowest-lying excited states (which
are $N\pi$ or $N\pi\pi$ states in a finite box) have smaller energy
gaps. Further adding to the difficulty is the requirement to compute a
new sequential propagator for each source-sink separation.

The problem of excited-state contamination has seen increased
attention in recent years, and alternatives to the ratio method have
been explored. These include the \emph{summation method}, where the
\emph{sums} of ratios are taken,
\begin{equation}
  S^\mathcal{O}(\vec p',\vec p,T) = \sum_{\tau=\tau_0}^{T-\tau_0} R^\mathcal{O}(\vec p',\vec p,\tau,T) = c + T\langle p'|\mathcal{O}|p\rangle + O(Te^{-\Delta E_\text{min}T}),
\end{equation}
where $\tau_0$ is a chosen parameter and $c$ is an unknown constant. The matrix element is then extracted from the slope of a line
fit to the sums at several values of $T$, or from a finite
difference. This yields improved asymptotic
behaviour~\cite{Capitani:2010sg,Bulava:2010ej}, with the leading
excited-state contaminants decaying as $Te^{-\Delta
  E_\text{min}T}$. Other approaches such as various forms of
multi-state
fits~\cite{Green:2011fg,Bhattacharya:2013ehc,Bali:2013nla,Bali:2014gha,Alexandrou:2014wca,vonHippel:2014hla,Junnarkar:2014jxa}
and the use of a variational basis of interpolating
operators~\cite{Aubin:2010jc,Owen:2012ts,Owen:2013pfa,Green:2014xba}
have also been explored.

Besides excited states, important systematics include the following:
\begin{enumerate}
\item Continuum extrapolation $a\to 0$. Depending on the operator and
  the discretization, effects may be $O(a)$ or $O(a^2)$. Evidence for
  a significant effect on nucleon observables has not been reported,
  however this could be an important issue.
\item Infinite-volume extrapolation $L_s\to\infty$. For hadronic
  matrix elements, the leading effects are $O(e^{-m_\pi L})$; the
  usual rule of thumb is that $m_\pi L > 4$ is sufficient, although
  there have been few careful studies.
\item Non-physical quark masses. Most calculations have been performed
  at heavier-than-physical pion masses and have relied on
  extrapolation (typically using some form of chiral perturbation
  theory) to the physical point. With advances in algorithms and
  computational power, some recent calculations have been performed
  near~\cite{Green:2012ud,Bali:2014gha} or
  at~\cite{Alexandrou:2014wca,Syritsyn_Lat14,Koutsou_Lat14,Gupta_Lat14}
  the physical pion mass. This is important for some observables that
  show strong dependence on the pion mass in the chiral regime, such
  as the charge radius, which diverges in the chiral limit.
\end{enumerate}

Calculations using physical pion masses have become much more
practical due to algorithmic advances, such as the related techniques
of truncated solver~\cite{Collins:2007mh,Bali:2009hu} and
all-mode-averaging (AMA)~\cite{Blum:2012uh}, which make use of a large
number of samples computed using approximate quark propagators,
supplemented with a relatively small number of samples computed using
exact quark propagators for bias correction. The approximation for the
quark propagators is chosen such that they can be computed much more
quickly and the contribution to the variance from bias correction is
small. These have proven to be essential tools, especially for
computationally-expensive actions such as domain wall
fermions~\cite{Lin:2014saa,Ohta:2014rfa,Syritsyn_Lat14}.

\section{Benchmark observables}

Given that full control over all systematics is still a work in
progress for nucleon structure calculations, we rely on comparisons
with experiment to help judge the quality of our
calculations. Understanding what is required to obtain agreement with
experiment for these ``benchmark'' observables, such as the axial
charge and electromagnetic form factors, is essential for judging the
quality of calculations of other observables.

\subsection{Axial charge}

The nucleon axial charge $g_A$ is defined via a neutron-to-proton
transition matrix element,
\begin{equation}
\langle p(P) | \bar u \gamma_5\gamma_\mu d | n(P)\rangle = g_A \bar u_p(P) \gamma_5\gamma_\mu u_n(P),
\end{equation}
and has long served as a benchmark for lattice calculations. It is a
relatively simple quantity to compute, being a forward matrix element
and an isovector quantity that doesn't require disconnected
diagrams. Experimentally, it is well known from beta decay of
polarized neutrons; the latest PDG value is
$g_A=1.2723(23)$~\cite{Agashe:2014kda}.

\begin{figure}
  \centering
  \includegraphics{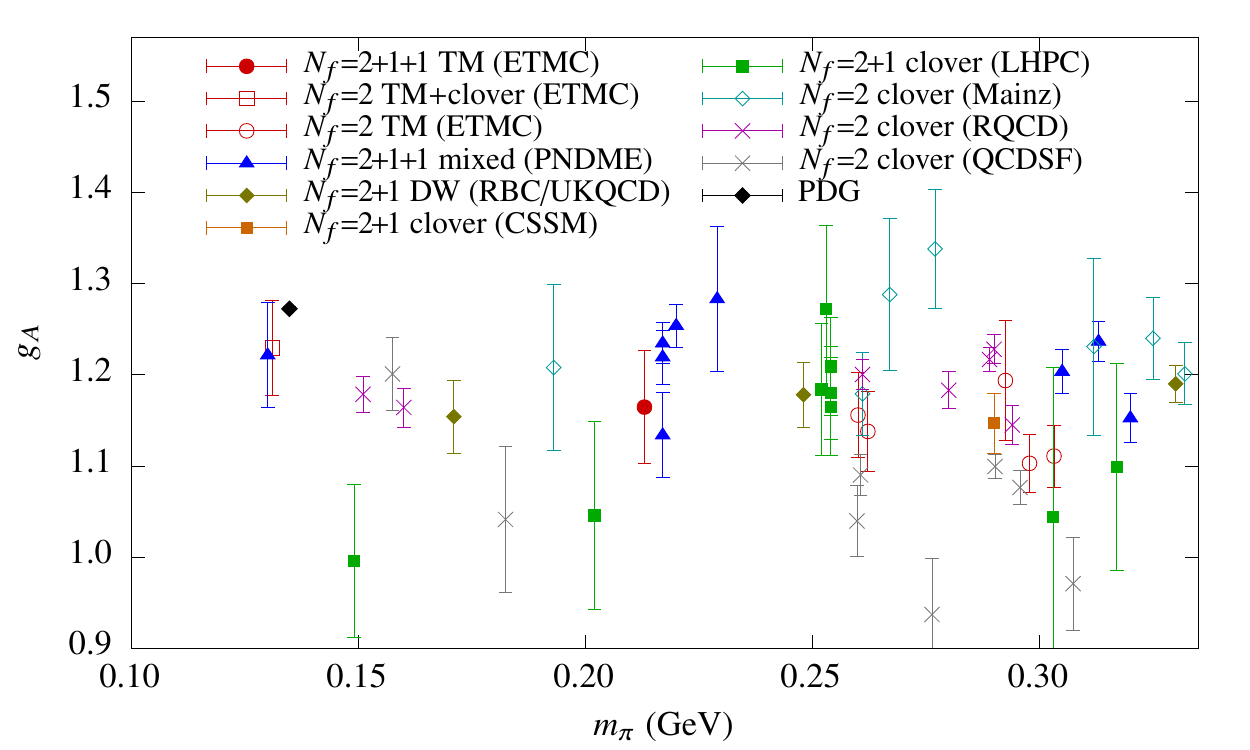}
  \caption{Axial charge versus pion mass, computed using $N_f=2$ and
    $N_f=2+1+1$ twisted mass fermions~\cite{Alexandrou:2014wca}; a
    mixed action with clover-improved Wilson valence quarks and
    $N_f=2+1+1$ HISQ staggered sea quarks~\cite{Gupta_Lat14};
    $N_f=2+1$ domain wall fermions~\cite{Ohta:2014rfa}; $N_f=2+1$
    clover-improved Wilson fermions, without~\cite{Owen:2012ts} and
    with~\cite{Green:2012ud} smearing; and $N_f=2$ clover-improved
    Wilson
    fermions~\cite{Jager:2013kha,Bali_Lat14,Horsley:2013ayv}. Note
    that some of the same ensembles were used by RQCD and QCDSF, so
    that their errors may be correlated. 
}
  \label{fig:gA_mpi}
\end{figure}

Obtaining agreement with the experimental value has proven difficult
for lattice calculations. Those that include pion masses below 300~MeV
are shown in Fig.~\ref{fig:gA_mpi}. Note that, unless otherwise
stated, the plotted data show the ``raw'' values from each lattice
ensemble, renormalized but without any extrapolations to zero lattice
spacing, infinite volume, or physical pion mass. In general, the
lattice data tend to lie below the experimental value and show no
strong dependence on the pion mass. In the past, the added uncertainty
due to extrapolations to the physical pion mass meant that data that
were below experiment could still be reconciled with
it~\cite{Edwards:2005ym}, but this becomes more difficult as
calculations with near-physical pion masses become available.

The possibility of large excited-state contaminations affecting
lattice calculations of $g_A$ has seen several studies in recent
years. The Mainz group obtained agreement of their extrapolated value
with experiment, when using the summation method to remove
contributions from excited states, whereas the ratio method with a
source-sink separation $T\approx 1.1$~fm produced a value below
experiment~\cite{Capitani:2012gj}. Using similar methods, LHPC
reported similar results for pion masses $m_\pi\gtrsim 250$~MeV, but
found that closer to the physical pion mass, removing excited states
yielded even lower values of $g_A$ than typical lattice
calculations~\cite{Green:2012ud}. Further evidence for the importance
of excited-state effects comes from comparing results from the RQCD
and QCDSF collaborations in Fig.~\ref{fig:gA_mpi}, where many of the
same lattice ensembles were used (thus controlling most systematics)
but different quark-field smearing was used in the interpolating
operator, leading to significantly different values of
$g_A$~\cite{Bali:2013nla}.

\begin{figure}
  \centering
  \includegraphics{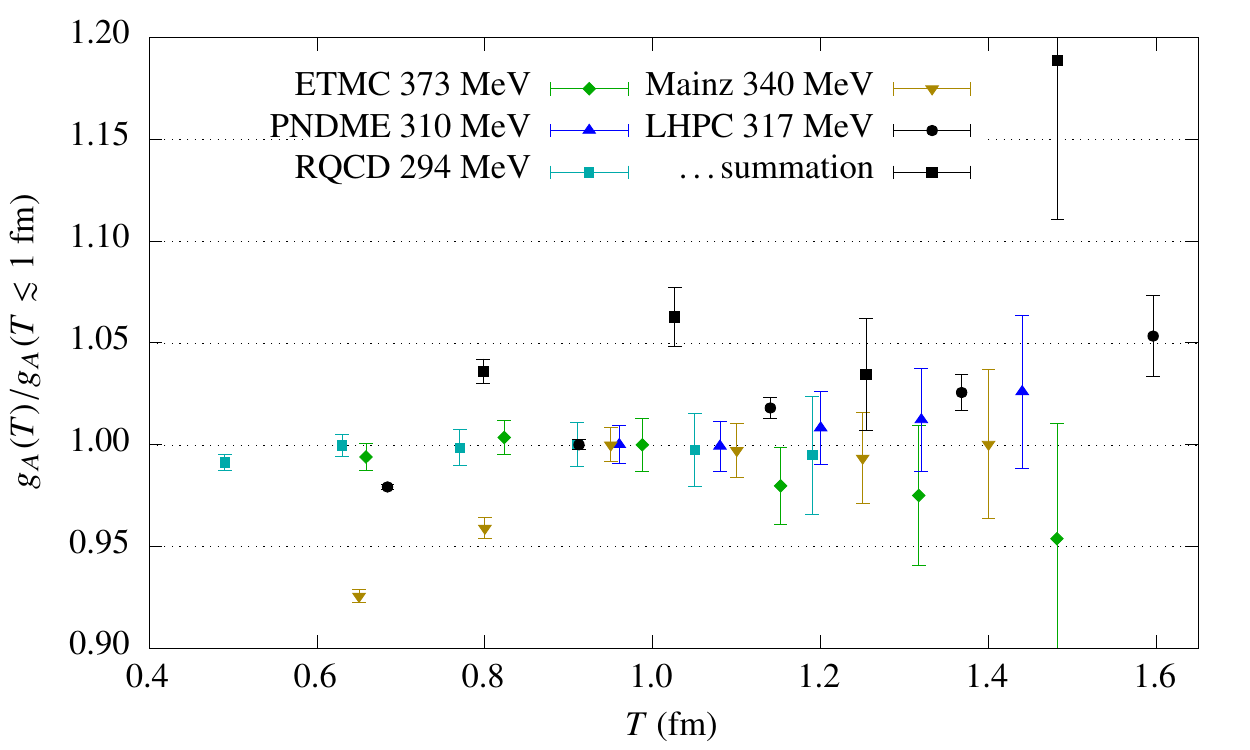}
  \caption{Axial charge, computed using the ratio method, versus
    source-sink separation $T$, with pion masses
    373~\cite{Alexandrou:2014wca}, 310~\cite{Bhattacharya:2013ehc},
    294~\cite{Bali:2013nla}, 340~\cite{Junnarkar:2014jxa}, and
    317~\cite{Green_thesis}~MeV. Data are normalized to their values
    at $T$ slightly below 1~fm. For the very-high-precision study
    from LHPC, we also show the summation-method results, computed
    using finite differences between the sums at the two nearest
    source-sink separations.}
  \label{fig:gA_tsink}
\end{figure}

Some dedicated studies of excited-state effects are shown in
Fig.~\ref{fig:gA_tsink}. The highest-precision data are from a
calculation by LHPC using very high statistics and a large volume;
these indicate a $-5\%$ shift caused by excited states, when using the
ratio method with a source-sink separation $T$ slightly below
1~fm. Other calculations have noisier results, but the dependence on
$T$ past 1~fm is consistent with the LHPC data.  On the other hand,
studying Fig.~\ref{fig:gA_tsink} below $T=1$~fm, where excited-state
effects are expected to be more prominent, shows different behaviour
among the different calculations. This may indicate that ETMC and RQCD
were more successful at tuning their interpolating operators to
eliminate these effects. Finally, this figure also shows the
effectiveness of the summation method at removing excited-state
effects using relatively small source-sink separations, albeit with
increased noise compared with the ratio method.

\begin{figure}
  \centering
  \includegraphics{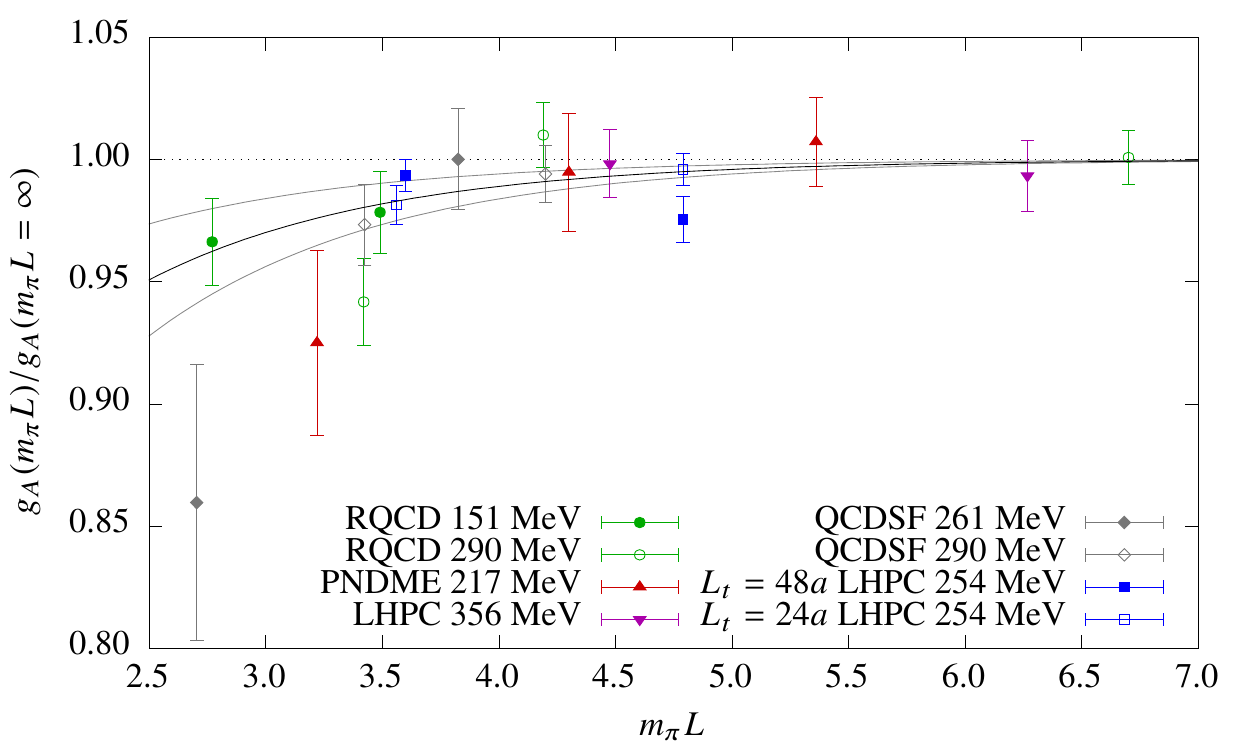}
  \caption{Axial charge versus the product of pion mass and spatial
    box length, $m_\pi L$, from RQCD~\cite{Bali_Lat14},
    QCDSF~\cite{Horsley:2013ayv}, PNDME~\cite{Gupta_Lat14}, and
    LHPC~\cite{Bratt:2010jn,Green:2013hja}. Data are normalized to
    their infinite-volume extrapolation (see text), and the black and
    gray curves show the fitted dependence on $m_\pi L$ and its
    $1\sigma$ error. Note that the two calculations with
    $m_\pi=290$~MeV used the same two ensembles, so a 50\% error
    correlation was assumed when fitting. At $m_\pi=254$~MeV, the
    spatial and temporal extents of the box were varied independently;
    here we treat the two temporal extents as two independent
    finite-volume studies.}
  \label{fig:gA_mpiL}
\end{figure}

The other major focus in studies of systematic errors for $g_A$ has
been on the finite-volume effects, which have been prominently
emphasized in works of the RBC-UKQCD collaboration, particularly in
Ref.~\cite{Yamazaki:2008py}, where $g_A$ was found to depend
significantly on $m_\pi L$, even in the region around $m_\pi L=4$, and
that $m_\pi L\gtrsim 6$ would be needed to keep finite-volume effects
below 1\%. There have been relatively few \emph{fully-controlled}
studies\footnote{Note that we exclude studies such as that of
  Ref.~\cite{Yamazaki:2008py} where the lattice time extent was
  changed together with its spatial extent.} of finite-volume
effects. These are shown in Fig.~\ref{fig:gA_mpiL}, where we see that
small volumes lead to small values of $g_A$. In order to study the
dependence on the box size, we fit the data using a floating norm with
a crude model,
\begin{equation}
  g_A(m_\pi L,\dots) = A(\dots)(1 + Be^{-m_\pi L}),
\end{equation}
where $A(\dots)$ depends on all parameters except for $m_\pi L$. I.e.,
the data in Fig.~\ref{fig:gA_mpiL} are fitted using eight independent
parameters $A(\dots)$ and a universal $B$ parametrizing the
finite-volume effects; then $A$ is used to normalize each dataset in
the figure. Although the model isn't a perfect description of the data
($\chi^2/\text{dof}=20/9$), it implies a small shift in the value of
$g_A$ by $-1.1(5)\%$ when $m_\pi L=4$.

It should also be noted that we are neglecting any interactions
between the box size and other systematic errors. For the case of
excited states, this could be an important effect if significant
contributions come from multi-particle $N\pi$ or $N\pi\pi$ states,
since the energy gaps and couplings to the interpolating operator
would depend on the box size. If we insist on controlling excited
states before studying finite-volume effects, then the constraints
that we can find on the latter are weaker; e.g.,
Ref.~\cite{Green:2013hja} indicates that $g_A$ is shifted by less than
5\% when $m_\pi L=4$.

At this point, there is no consensus, among those who perform lattice
QCD calculations of nucleon structure, regarding a single culprit for
the long-standing discrepancy with experimental measurements of
$g_A$. The very recent ability to perform calculations at the physical
pion mass will eliminate one source of uncertainty, but it seems that
the axial charge will remain a troublesome observable for the near
future. The problems may ultimately prove to come from a combination
of multiple systematic errors, and it is possible that less-studied
issues may be important, such as thermal states arising from a finite
time-extent~\cite{Green:2012ud} or inefficient sampling of gauge
fields leading to a long-range autocorrelation~\cite{Ohta:2014rfa}.

\subsection{Electromagnetic form factors}

The Dirac and Pauli form factors of the vector current,
\begin{equation}
  \langle p'|\bar q\gamma^\mu q|p\rangle = \bar u(p')\left(\gamma^\mu F_1^q(Q^2) + \frac{i\sigma^{\mu\nu}(p'-p)_\nu}{2m_N} F_2^q(Q^2) \right)u(p),
\end{equation}
where $Q^2=-(p'-p)^2$, have been the primary off-forward benchmarks
for lattice nucleon-structure calculations. The isovector ($u-d$)
combination can be compared with the difference between proton and
neutron form factors from elastic scattering with electrons, which are
usually given in terms of electric and magnetic Sachs form factors,
\begin{align}
  G_E(Q^2) &= F_1(Q^2) - \frac{Q^2}{(2m_N)^2}F_2(Q^2), \\
  G_M(Q^2) &= F_1(Q^2) + F_2(Q^2).
\end{align}

As was the case for the axial charge, lattice calculations of the
electromagnetic form factors have disagreed with experiment, producing
a much milder dependence on $Q^2$. For a long time this was easily
attributable to the use of heavier-than-physical pion masses, since
chiral perturbation theory predicts that both the isovector charge and
magnetic radii diverge in the chiral limit. However, this discrepancy
persisted as pion masses were reduced.

\begin{figure}
  \centering
  \includegraphics[width=0.49\textwidth]{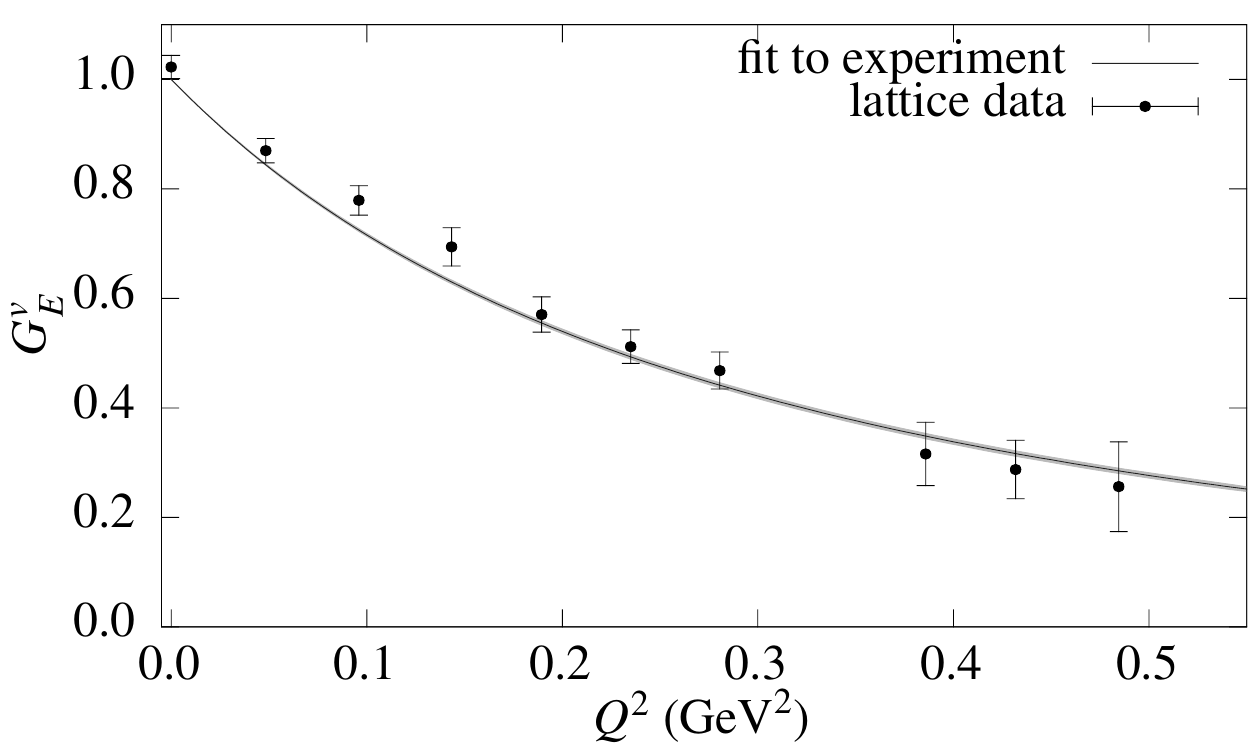}
  \includegraphics[width=0.49\textwidth]{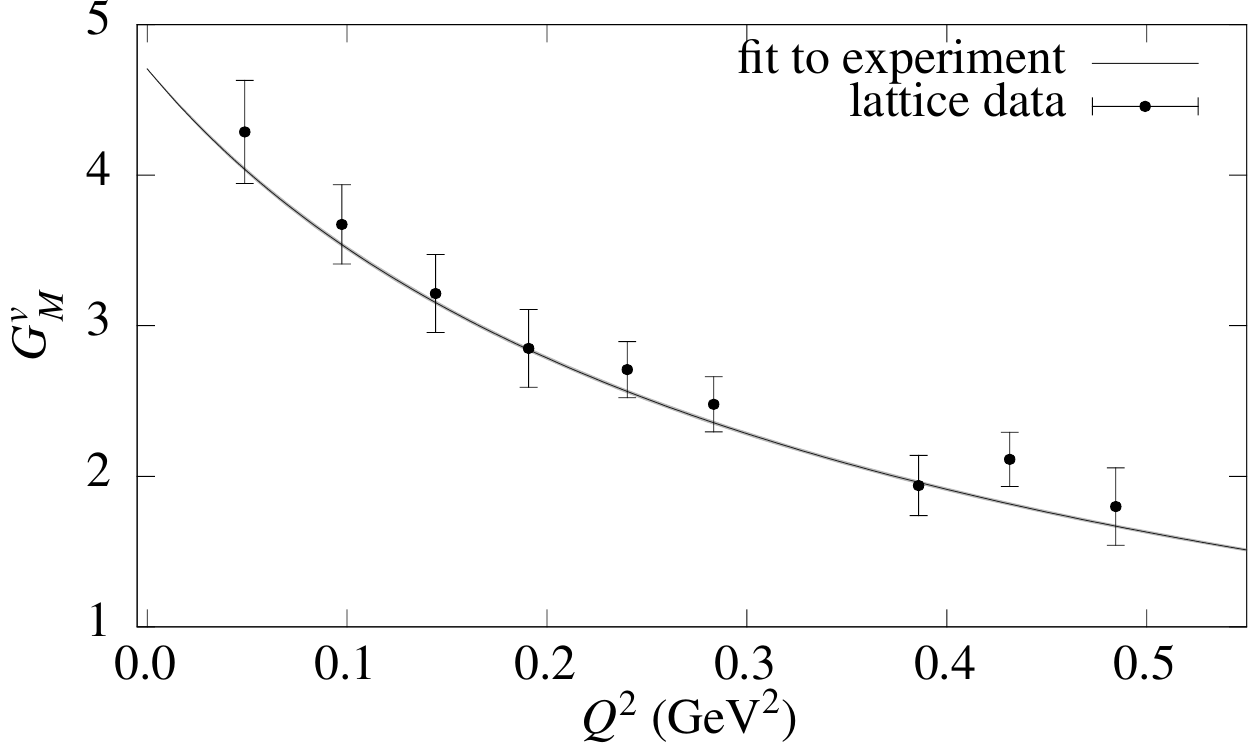}
  \caption{Isovector electromagnetic form factors. The lattice data
    are from the calculation in
    Refs.~\cite{Green:2013hja,Green:2014xba}, with $m_\pi=149$~MeV and
    using the summation method to reduce excited-state effects. The
    curves (including narrow error bands) are from the fit to
    experimental data in Ref.~\cite{Alberico:2008sz}.}
  \label{fig:GE_GM_expt}
\end{figure}

Studies of contributions from excited states have found large
effects~\cite{Green:2011fg,Green:2014xba,vonHippel:2014hla,Koutsou_Lat14},
and that by reducing them and using relatively light pion masses, much
of the gap between lattice calculations and experiment could be
closed. This is displayed in Fig.~\ref{fig:GE_GM_expt}, which shows
results from a calculation where the isovector $G_E$ and $G_M$ agree
with experiment~\cite{Green:2013hja,Green:2014xba}. Although this
needs to be reproduced and successful control over excited states
remains a significant challenge, these results are an encouraging sign
of progress in nucleon structure calculations.

The behaviour of the form factors near $Q^2=0$ is of particular
interest, as it yields the charge and magnetic radii ($r_E^2$ and
$r_M^2$) and the magnetic moment $\mu$,
\begin{align}
  G_E(Q^2) &= 1 - \frac{Q^2}{6} r_E^2 + O(Q^4),\\
  G_M(Q^2) &= \mu\left(1 - \frac{Q^2}{6} r_M^2 + O(Q^4) \right).
\end{align}
In particular, experimental determinations of the proton charge radius
have a 7$\sigma$ discrepancy between the very precise value determined
from spectroscopy of muonic hydrogen~\cite{Antognini:1900ns},
$r_E^p=0.84087(39)$~fm, and the CODATA value determined from
scattering and spectroscopy with electrons~\cite{Mohr:2012tt},
$r_E^p=0.8775(51)$~fm. Having a reliable \emph{ab initio} calculation
of the proton charge radius (or even the isovector charge radius) is
thus a highly attractive goal for practitioners of lattice QCD. Since
the discrepancy in the squared proton charge radius $(r_E^2)^p$ is
8--9\%, even though lattice calculations are uncompetitive with the
experimental precision, distinguishing between the two experimental
values may be within reach in the next few years.

Given a form factor computed on the lattice, fitting is required to
determine the radii and magnetic moment. This is usually done with a
two-parameter dipole fit,
\begin{equation}
  F(Q^2) = \frac{F(0)}{\left(1+\frac{Q^2}{m_D^2}\right)^2},
\end{equation}
although other forms have also been explored. Once form factors are
reliably computed at the physical pion mass, this model-dependent
fitting will probably be a leading source of uncertainty in radii and
the magnetic moment. The situation could be improved with data at
smaller $Q^2$; however, with periodic boundary conditions, the
smallest available nonzero spatial momentum transfer is $2\pi/L_s$, so
that even with the large $(5.6\text{ fm})^3$ volume used for the
calculation shown in Fig.~\ref{fig:GE_GM_expt}, the minimum momentum
transfer was $Q^2=0.05\text{ GeV}^2$, an order of magnitude larger
than obtained in scattering
experiments~\cite{Murphy:1974zz,Simon:1980hu,Bernauer:2013tpr}.

We describe four strategies that might be used to obtain a better
calculation of radii:
\begin{enumerate}
\item Use larger volumes to reduce $2\pi/L$ and thus the minimum
  $Q^2$. This is the most straightforward approach, and it is
  compatible with disconnected diagrams, so that the proton form
  factors could be directly computed. To access very low $Q^2$ without
  a rapid increase in computational costs, asymmetric boxes could be
  used, with one ``long'' spatial dimension.
\item Boost the source and sink nucleons to large momentum, in the
  direction of momentum transfer. This way, small $Q^2$ can be reached
  by increasing the energy difference between the source and the sink
  while keeping the spatial momentum transfer fixed to its minimum
  value. This strategy would likely require a nonstandard approach to
  the interpolating operators, since the usual smeared operators tend
  to have a poor signal and poor overlap with the ground state at
  large momentum~\cite{Lin:2010fv,Roberts:2012tp,DellaMorte:2012xc}.
\item Use (partially) twisted boundary
  conditions~\cite{Bedaque:2004kc,deDivitiis:2004kq,Sachrajda:2004mi}. Using
  different boundary conditions on different quark flavors allow for
  arbitrary momentum transfer to be probed. The \emph{partially}
  twisted case corresponds to only changing the boundary conditions on
  the valence quarks and not on the sea quarks, in order to reuse an
  existing ensemble of gauge configurations. However, twisted boundary
  conditions can introduce additional finite-volume effects that are
  exponentially suppressed at large volume, but may be significant in
  practice, especially for
  $F_2(Q^2)$~\cite{Jiang:2008ja,Tiburzi:2006px}. Since this technique
  relies on a transition between two flavors with different boundary
  conditions, it cannot be applied to disconnected diagrams, meaning
  that a clean, direct calculation of the proton form factors is not
  possible. There has been one preliminary study of the application
  of partially twisted boundary conditions to nucleon isovector form
  factors~\cite{Gockeler:2008zz}.
\item Use the Rome method~\cite{deDivitiis:2012vs}. This amounts to
  setting up a calculation using partially twisted boundary
  conditions, and then analytically taking the derivative with respect
  to the twist angle, which requires computing additional sequential
  propagators. By evaluating the derivative at zero twist angle,
  momentum-derivatives of matrix elements at $Q^2=0$ can be computed,
  leading to a direct computation of the radii, independent of any
  fitting to $F(Q^2)$. This has been studied for the pion in chiral
  perturbation theory~\cite{Tiburzi:2014yra}, where it was found that
  finite-volume effects for the charge radius asymptotically scale as
  $L^{1/2}e^{-m_\pi L}$. As with twisted boundary conditions, this
  method cannot be applied to disconnected diagrams.
\end{enumerate}
Together with good control over excited states and the use of physical
quark masses, these techniques may help to produce a reliable
QCD calculation of the proton charge radius.

\section{Other observables}

%...

\subsection{Momentum fraction}

The quark and gluon momentum fractions in a proton are obtained from
forward matrix elements of the traceless energy-momentum tensor,
\begin{equation}
  \langle p| T^{\mu\nu}_{q,g} |p\rangle = \langle x\rangle_{q,g} \bar u(p) \gamma^{\{\mu}p^{\nu\}} u(p),
\end{equation}
where the braces denote taking the traceless symmetric part and
\begin{align}
  T^{\mu\nu}_q &= \bar q \gamma^{\{\mu}i\overleftrightarrow{D}^{\nu\}}q,\\
  T^{\mu\nu}_g &= G^{\{\mu\alpha a}{G_\alpha}^{\nu\}a}.
\end{align}
These satisfy a sum rule: $\langle x\rangle_g + \sum_q\langle x\rangle_q=1$.

\begin{figure}
  \centering
  \includegraphics{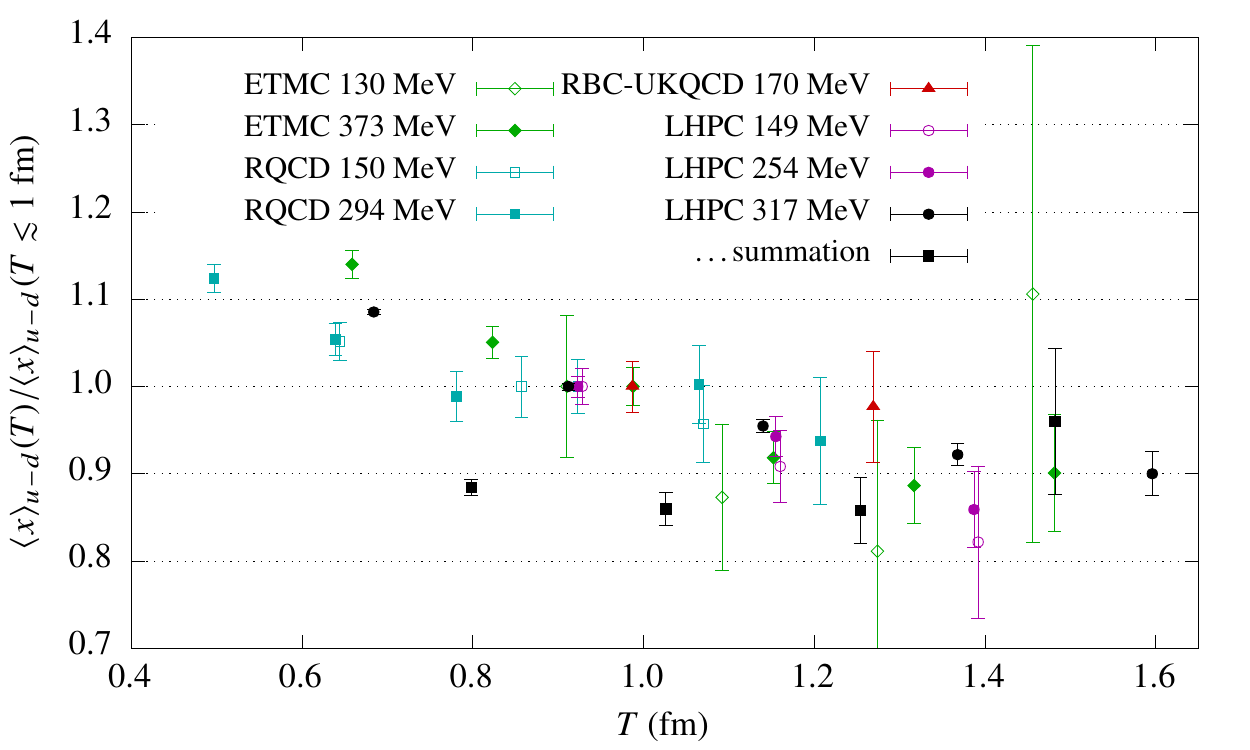}
  \caption{Isovector average quark momentum fraction, computed using
    the ratio method, versus source-sink separation $T$, with pion
    masses 130 and 373~\cite{Alexandrou:2014wca}, 150 and
    294~\cite{Bali:2014gha}, 170~\cite{Ohta:2013qda}, 149 and
    254~\cite{Green:2012ud}, and 317~\cite{Green_thesis}~MeV. Data are
    normalized to their values at $T$ slightly below 1~fm. For the
    very-high-precision study from LHPC, we also show the
    summation-method results, computed using finite differences
    between the sums at the two nearest source-sink separations.}
  \label{fig:x_tsink}
\end{figure}

In order to avoid the need for disconnected diagrams, most
calculations have been focussed on the isovector combination, $\langle
x\rangle_{u-d}$. It has been found that this observable suffers from
large excited-state effects; a selection of excited-state studies is
shown in Fig.~\ref{fig:x_tsink}. These are broadly consistent with one
another and indicate that using the ratio method with source-sink
separation slightly below 1~fm yields a value 10--15\% above the
ground-state value, although the effect may grow at smaller pion
masses~\cite{Green:2012ud}. It seems clear that good control over
excited states is essential for the momentum fraction; indeed, the
calculation of Ref.~\cite{Green:2012ud}, using the summation method to
reduce their effect and extrapolating to the physical pion mass, found
$\langle x\rangle_{u-d}=0.140(21)$ (in the $\overline{\text{MS}}$
scheme at 2~GeV), which is in agreement with the phenomenological
value.

Going beyond the isovector combination, we note the study by $\chi$QCD
in which the gluon and individual $u$, $d$, and $s$ quark
contributions to the nucleon's momentum were
computed~\cite{Deka:2013zha}, albeit using fairly heavy quark masses
and a quenched ensemble. There also have been some other recent
studies of $\langle x\rangle_g$, using quenched~\cite{Horsley:2012pz}
and dynamical~\cite{Alexandrou:2013tfa} fermions, and it is
encouraging that reasonable statistical errors can be obtained. Some
studies of the disconnected contributions to the quark momentum
fraction will be briefly discussed in the later section on
disconnected diagrams. These observables also face a more complicated
pattern of renormalization, where the quark and gluon momentum
fractions mix. The mixing coefficients have been studied
perturbatively~\cite{Glatzmaier:2014sya} but not yet
nonperturbatively.

\subsection{Scalar and tensor charges}

The nucleon scalar charge $g_S$ and tensor charge $g_T$ are defined,
analogously to the axial charge, via neutron-to-proton transition
matrix elements:
\begin{align}
\langle p(P) | \bar u d | n(P)\rangle &= g_S \bar u_p(P) u_n(P),\\
\langle p(P) | \bar u \sigma^{\mu\nu} d | n(P)\rangle &= g_T \bar u_p(P) \sigma^{\mu\nu} u_n(P).
\end{align}
In a study of generic beyond-the-Standard-Model contributions to
neutron beta decay, it was shown that the leading effects are
proportional to these two couplings; thus, calculations of $g_S$ and
$g_T$ are required in order to find constraints on BSM physics from
beta-decay experiments~\cite{Bhattacharya:2011qm}.

The tensor charge is also equal to the isovector first moment of the
proton's transversity,
\begin{equation}
  g_T = \langle 1\rangle_{\delta u- \delta d}.
\end{equation}
Experimental measurements with good precision are planned at Jefferson
Lab~\cite{Dudek:2012vr}; therefore the tensor charge is an attractive
target for testing predictions of lattice QCD.

The scalar charge is related via the Feynman-Hellmann theorem to the
contribution from the difference in $u$ and $d$ quark masses to the
neutron-proton mass splitting,
\begin{equation}
  g_S = \frac{(m_n-m_p)_\text{QCD}}{m_d-m_u}\label{eq:gS_FH}
\end{equation}
(up to higher order isospin-breaking corrections), which also provides
an indirect way of calculating the scalar charge using lattice
QCD~\cite{Gonzalez-Alonso:2013ura}.

\begin{figure}
  \centering
  \includegraphics{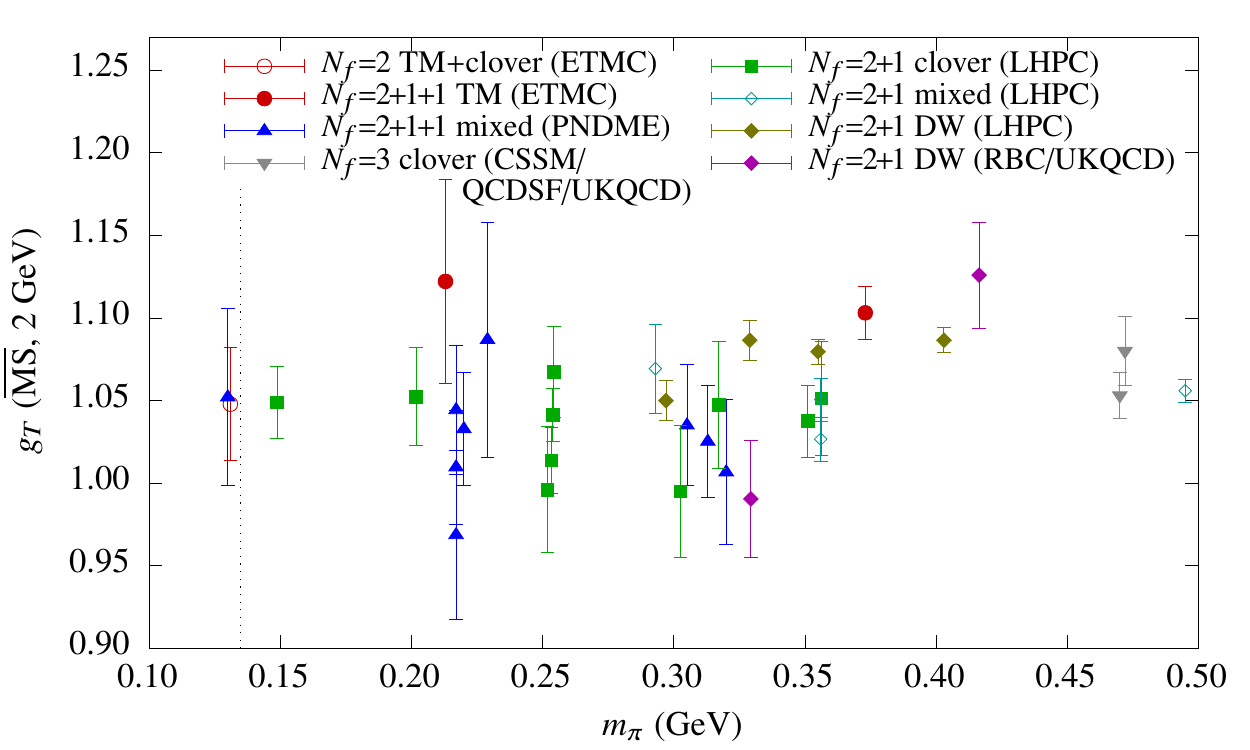}
  \caption{Nucleon tensor charge (renormalized in the
    $\overline{\text{MS}}$ scheme at 2~GeV) versus pion mass, computed
    using $N_f=2$ and $N_f=2+1+1$ twisted mass
    fermions~\cite{Alexandrou:2014wca}, a mixed action with
    clover-improved Wilson valence quarks and $N_f=2+1+1$ HISQ
    staggered sea quarks~\cite{Gupta_Lat14},
    $N_f=3$~\cite{Zanotti_Lat14} and
    $N_f=2+1$~\cite{Green:2012ej,Green:2012rr} clover-improved Wilson
    fermions with smearing, a mixed action with domain wall valence
    quarks and $N_f=2+1$ Asqtad staggered sea
    quarks~\cite{Green:2012ej}, and $N_f=2+1$ domain wall
    fermions~\cite{Green:2012ej,Aoki:2010xg}. The vertical dotted line
    indicates the physical pion mass.}
  \label{fig:gT_mpi}
\end{figure}

Lattice QCD calculations of the tensor charge are summarized in
Fig.~\ref{fig:gT_mpi}. There is no sign of a significant dependence on
the pion mass. In addition, when excited-state effects have been
studied for $g_T$, they have been found to be
small~\cite{Green:2012ej,Bhattacharya:2013ehc,Alexandrou:2014wca}, and
the preliminary extrapolation in Ref.~\cite{Gupta_Lat14} to the
continuum and infinite-volume limits found only mild effects. Thus the
tensor charge appears to be a well-behaved observable, and should
serve as a good test for lattice calculations when precise
experimental measurements become available.

\begin{figure}
  \centering
  \includegraphics{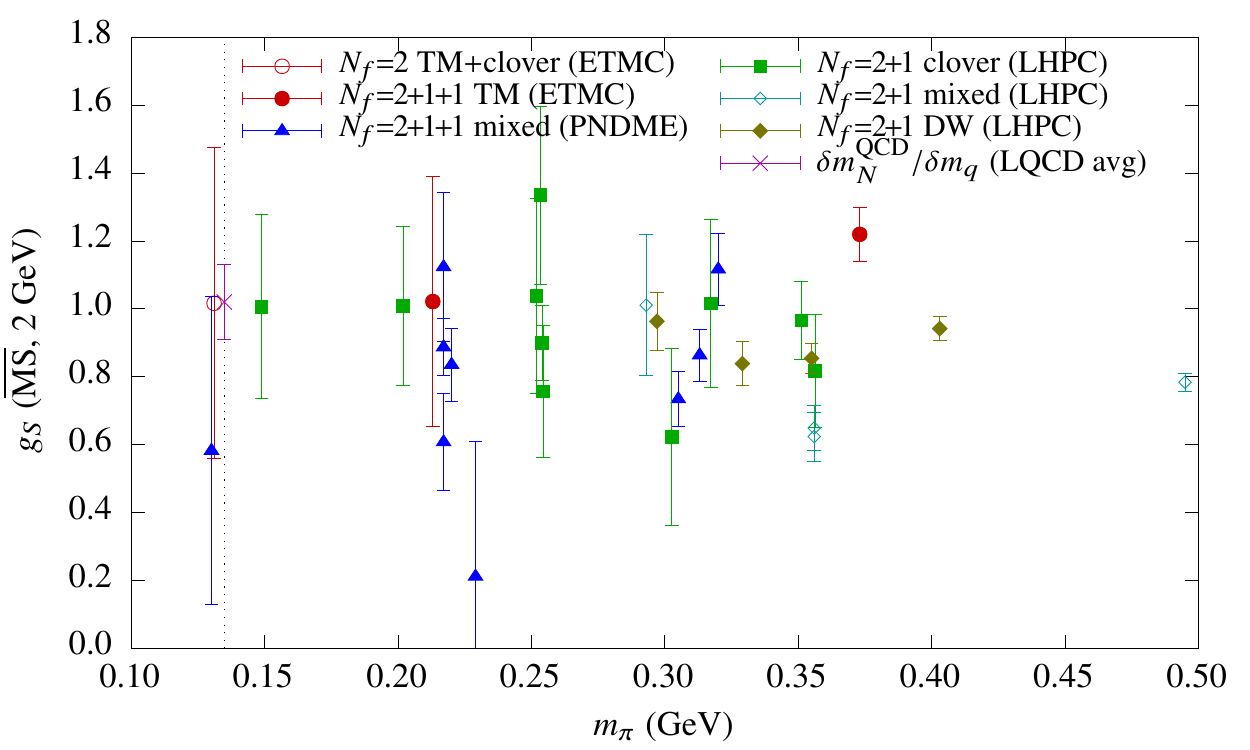}
  \caption{Nucleon scalar charge (renormalized in the
    $\overline{\text{MS}}$ scheme at 2~GeV) versus pion mass, computed
    using $N_f=2$ and $N_f=2+1+1$ twisted mass
    fermions~\cite{Alexandrou:2014wca}, a mixed action with
    clover-improved Wilson valence quarks and $N_f=2+1+1$ HISQ
    staggered sea quarks~\cite{Gupta_Lat14}, $N_f=2+1$ clover-improved
    Wilson fermions with smearing~\cite{Green:2012ej,Green:2012rr}, a
    mixed action with domain wall valence quarks and $N_f=2+1$ Asqtad
    staggered sea quarks~\cite{Green:2012ej}, $N_f=2+1$ domain wall
    fermions~\cite{Green:2012ej}, and indirectly using
    Eq.~(\ref{eq:gS_FH}) with lattice QCD data at the physical
    point~\cite{Gonzalez-Alonso:2013ura}. The vertical dotted line
    indicates the physical pion mass.}
  \label{fig:gS_mpi}
\end{figure}

Similarly, we show the scalar charge in Fig.~\ref{fig:gS_mpi}. This
suffers from considerably more noise than the axial and tensor
charges; as a result, it is difficult to clearly identify systematic
effects. ETMC reports large excited-state
effects~\cite{Alexandrou:2014wca}, but higher statistics are needed to
confirm that this is a problem.

\section{Disconnected diagrams}

The disconnected-diagrams contributions to a hadron three-point
function equal the correlation between a \emph{disconnected loop},
\begin{equation}
  T(\vec p'-\vec p,\tau) = \sum_{\vec y}e^{i(\vec p'-\vec p)\cdot(\vec y-\vec x_0)}\mathcal{O}(y,t_0+\tau),
\end{equation}
and the hadron's two-point function, Eq.~(\ref{eq:c2pt}). For a local
quark bilinear operator, $\mathcal{O}=\bar q\Gamma q$, this requires
computing the quark propagator from every point $y$ back to itself,
$\Tr(\Gamma S[U](y,y))$, which is typically done using stochastic
estimation by introducing noise vectors $\eta$ that have expectation
value $E(\eta \eta^\dagger)=I$. Solving $D[U]\psi=\eta$ yields a
stochastic estimate for the all-to-all propagator,
$S[U]=E(\psi\eta^\dagger)$. Various techniques are used to reduce the
noise associated with this procedure more efficiently than the
$1/\sqrt{n}$ obtained by using many noise sources.

\begin{table}
  \centering
  \begin{tabular}{lrrr}
    & ETMC~\cite{Abdel-Rehim:2013wlz,Alexandrou:2014axa} 
    & LHPC~\cite{Meinel_Lat14}
    & LANL~\cite{Yoon_Lat14} \\\hline
    $N_f$ & $2+1+1$ & $2+1$ & $2+1+1$\\
    Action & twisted mass & clover & clover on HISQ \\
    $a$ & 0.082~fm & 0.114~fm & 0.12~fm \\
    $m_\pi$ & 372~MeV & 317~MeV & 305~MeV \\
    $N_\text{samp}(C_\text{2pt})$ & 147k & 99k & 61k\\
    Methods & truncated solver~\cite{Collins:2007mh,Bali:2009hu}, & hierarchical probing~\cite{Stathopoulos:2013aci} & truncated solver, \\
    ~~for loops & one-end trick~\cite{Michael:2007vn} &                                               & hopping parameter expansion~\cite{Collins:2007mh,Bali:2009hu} \\
    \hline
    $R(g_A^{u+d})$ & $-0.12(2)$ & $-0.12(2)$ & $-0.19(2)$\\
    $R(g_T^{u+d})$ & $-0.002(2)$ & $-0.005(10)$ & $-0.039(8)$\\
    $R(g_S^{u+d})$ & $0.101(15)$ & $1.756(94)$ & $0.328(25)$\\
    $R(\langle x\rangle_{u+d})$ & $0.05(13)$ & $0.24(4)$ & ---~~~
  \end{tabular}
  \caption{Recent calculations of light quark disconnected contributions to isoscalar nucleon forward matrix elements. $R(X)$ is defined as the ratio of disconnected to connected contributions to the observable $X$.}
  \label{tab:disc_diag}
\end{table}

There are an increasing number of nucleon-structure calculations that
include disconnected diagrams, however these generally do not cover
the same range of ensembles as for connected diagrams, so that less is
known about systematic errors. Three such calculations that were
presented at the Lattice 2014 conference, covering a range of
observables, are shown in Tab.~\ref{tab:disc_diag}. These used
$O(10^5)$ samples of the two-point function in order to obtain a good
signal (cf.\ typically $10^3$--$10^4$ samples for connected
three-point functions), along with various stochastic-estimation
techniques for the disconnected loops.  The importance of disconnected
contributions is highly observable-dependent: they form a moderate
positive contribution to the isoscalar scalar charge (relevant for the
nucleon sigma term) and momentum fraction, a moderate negative
contribution to the isoscalar axial charge, and a small contribution
to the isoscalar tensor charge.

For the light-quark disconnected contribution to electromagnetic form
factors, a clear nonzero signal was found by
LHPC~\cite{Meinel_Lat14}. For the proton $G_E$ and $G_M$, these were
less than 1\% of the connected part (positive for $G_E$, negative for
$G_M$). This calculation made use of hierarchical probing, an approach
that eliminates the variance of $\eta(x)\eta^\dagger(y)$ for nearby
spatial sites $x$ and $y$, and was found to be particularly effective
at reducing the noise for the vector current.

These calculations are still in their infancy for most nucleon
observables: the pion mass dependence and other systematics have not
been probed. But these first calculations are a promising sign that
reasonable signals can be obtained using existing techniques.

\section{Conclusions}

Lattice QCD calculations of nucleon structure have been making steady
progress toward full control over systematic errors. The ability to
calculate using the physical pion mass will nearly eliminate the
uncertainty associated with chiral extrapolation. Excited-state
contamination has been identified as an important source of errors,
and due to the exponentially-decaying signal, removing it remains a
challenge.

The axial charge remains the primary ``benchmark'' observable for
nucleon structure calculations, and agreement with experiment remains
problematic --- even in the cases where agreement within errors was
obtained, the lattice values are mostly below the experimental value,
rather than scattered on both sides of it. Until this issue is
unambiguously resolved, predictions of other observables will likely
have to be treated with some caution.

For electromagnetic form factors, it appears that the approach to the
physical pion mass and the removal of excited-state effects are the
most important systematics to control. It is promising that a
calculation has produced agreement with the experimental form factors,
and that a good signal has been obtained for the disconnected
contribution. The application of new techniques to probe the form
factors at low $Q^2$ may help to produce a solid first-principles
calculation of the proton charge radius with sufficient accuracy to
discriminate between the two experimental values.

The same methodology is being applied to observables such as the quark
and gluon momentum fractions, and the scalar and tensor charges. The
sources of uncertainty vary significantly: excited states are
significant for the momentum fraction, whereas statistical
fluctuations are quite large for the scalar charge. The tensor charge
appears to be under reasonably good control, and the agreement among
lattice calculations that its value is around 1.05 will provide a good
test once it is measured experimentally.

\section{Acknowledgments}

I thank John Negele for comments on a draft of this review as well as
my other collaborators in LHPC for their invaluable contributions to
some of the work presented here: Michael Engelhardt, Stefan Krieg,
Stefan Meinel, Andrew Pochinsky, and Sergey Syritsyn. I am also
grateful to my colleagues at Mainz for their helpful comments on an
early version of this talk.

\bibliographystyle{aipproc}   % if natbib is available

\bibliography{hadronstructure}

\end{document}